\documentclass[a4paper,12pt]{article}
\usepackage{graphics}

\begin{document}

\title{Spectra of radiation and created particles at intermediate
energy in oriented crystal taking into account energy loss}
\author{V. N. Baier
and V. M. Katkov\\
Budker Institute of Nuclear Physics,\\ Novosibirsk, 630090, Russia}

\maketitle

\begin{abstract}
The spectral distribution of positron created by photon and the
spectral distribution of photons radiated from electron in an
oriented single crystal of intermediate thickness is calculated at
intermediate energies. The energy loss of charged particles as well
as photon absorption are taken into account. The used basic
probabilities of processes include the action of field of axis as
well as the multiple scattering of radiating electron or particles
of the created pair (the Landau-Pomeranchuk-Migdal (LPM) effect).
\end{abstract}

\newpage

\section{Introduction}

1.In crystals oriented along main axes the probabilities of photon
emission from an electron and $e^+e^-$ pair creation by a photon are
strongly enhanced comparing to the corresponding amorphous media.
The example is shown in Fig.1,  where the crystal radiation length
$L(\varepsilon)=\varepsilon/I(\varepsilon),~ I(\varepsilon)$ is the
intensity of electron radiation, and the pair creation length
$L_{pr}(\omega)=1/W(\omega),~W(\omega)$ is the pair creation
probability, are plotted as a function of the corresponding energy.
The both functions $I(\varepsilon)$ and $W(\omega)$ are calculated
in frame of the new method, developed recently by authors
\cite{BK0,BK1}, which permits inseparable consideration of both
coherent and incoherent mechanisms of the process. It is seen that
there are two regions of $L_{pr}(\omega)$. In the first one
$L_{pr}(\omega)$ is almost constant. This is the incoherent
contribution (or, in other words, the Bethe-Maximon contribution
with the crystal corrections). It is seen that in this region $L \ll
L_{pr}$, what means that the probability of pair creation process is
still close to the probability in the corresponding amorphous medium
while the radiation length $L(\varepsilon)$ (connected with the
electron energy loss) is strongly enhanced (mostly due to the soft
photons emission) comparing with the corresponding amorphous medium
and the energy loss of electrons and positrons is strongly
influenced on the processes. This is true in the first region where
to one can add the vicinity of the turn point of the function
$L_{pr}(\omega)$ at $\omega \simeq \omega_m \equiv \varepsilon_m$
(see Table) where $L(\varepsilon_m)W(\omega_m) \ll 1$, e.g. in
tungsten at $\varepsilon=20~$GeV $ L_{pr} \simeq 4.9 L$ and in
germanium at $\varepsilon=200~$GeV $ L_{pr} \simeq 5.5 L$. In this
(intermediate) electron and photon energy  region it is impossible
to separate the coherent and incoherent contributions to the
probability of basic processes. In particular, one can't represent
the probability as the sum of contribution in the crystal field and
the Bethe-Maximon contribution. This is because one has to include
action of the crystal field on the incoherent process as well as the
multiple scattering of emitting electron or particles of created
pair for the process in field. These items were discussed in the
authors recent paper \cite{BK4}.

2.Basing on Eqs.(16) and (17) of \cite{BK0} (see also Eq.(7.135) in
\cite{BKS}) one get the general expression for the spectral
distribution of particles created by a photon
\begin{eqnarray}
&& dW(\omega, y)=\frac{\alpha m^2}{2\pi \omega} \frac{dy}{y(1-y)}
\int_0^{x_0}\frac{dx}{x_0}G(x, y),\quad G(x, y)=\int_0^{\infty} F(x,
y, t)dt +s_3\frac{\pi}{4},
\nonumber \\
&& F(x, y, t)={\rm Im}\left\lbrace e^{f_1(t)}\left[s_2\nu_0^2
(1+ib)f_2(t)-s_3f_3(t) \right] \right\rbrace,\quad
b=\frac{4\kappa_1^2}{\nu_0^2}, \quad y=\frac{\varepsilon}{\omega},
\nonumber \\
&& f_1(t)=(i-1)t+b(1+i)(f_2(t)-t),\quad
f_2(t)=\frac{\sqrt{2}}{\nu_0}\tanh\frac{\nu_0t}{\sqrt{2}},
\nonumber \\
&&f_3(t)=\frac{\sqrt{2}\nu_0}{\sinh(\sqrt{2}\nu_0t)}, \label{2}
\end{eqnarray}
where
\begin{equation}
s_2=y^2+(1-y)^2,~s_3=2y(1-y),~\nu_0^2=4y(1-y)
\frac{\omega}{\omega_c(x)},~\kappa_1=y(1-y)\kappa(x), \label{3}
\end{equation}
$\varepsilon$ is the energy of one of the created particles.

The situation is considered when the photon angle of incidence
$\vartheta_0$ (the angle between photon momentum {\bf k} and the
axis) is small under condition $\vartheta_0 \ll V_0/m$. The axis
potential (see Eq.(9.13) in \cite{BKS}) is taken in the form
\begin{equation}
U(x)=V_0\left[\ln\left(1+\frac{1}{x+\eta} \right)-
\ln\left(1+\frac{1}{x_0+\eta} \right) \right], \label{4}
\end{equation}
where
\begin{equation}
x_0=\frac{1}{\pi d n_a a_s^2}, \quad  \eta_1=\frac{2
u_1^2}{a_s^2},\quad x=\frac{\varrho^2}{a_s^2}, \label{5}
\end{equation}
Here $\varrho$ is the distance from axis, $u_1$ is the amplitude of
thermal vibration, $d$ is the mean distance between atoms forming
the axis, $a_s$ is the effective screening radius of the potential.
The parameters in Eq.(\ref{4}) were determined by means of fitting
procedure, see Table.

The local value of parameter $\kappa(x)$  which determines the
probability of pair creation in the field Eq.(\ref{4}) is
\begin{equation}
\kappa(x)=-\frac{dU(\varrho)}{d\varrho}\frac{\omega}{m^3}=2\kappa_sf(x),\quad
f(x)=\frac{\sqrt{x}}{(x+\eta)(x+\eta+1)},\quad \kappa_s=\frac{V_0
\omega}{m^3a_s}\equiv \frac{\omega}{\omega_s}.
 \label{6}
\end{equation}
For an axial orientation of crystal the ratio of the atom density
$n(\varrho)$ in the vicinity of an axis to the mean atom density
$n_a$ is (see \cite{BK0})
\begin{equation}
\frac{n(x)}{n_a}=\xi(x)=\frac{x_0}{\eta_1}e^{-x/\eta_1},\quad
\omega_0=\frac{\omega_e}{\xi(0)}, \quad
\omega_e=4\varepsilon_e=\frac{m}{4\pi
Z^2\alpha^2\lambda_c^3n_aL_0}.\label{7}
\end{equation}

The functions and values in Eqs.(\ref{2}) and (\ref{3}) are
\begin{eqnarray}
&&\omega_c(x)=
\frac{\omega_e(n_a)}{\xi(x)g_p(x)}=\frac{\omega_0}{g_p(x)}e^{x/\eta_1},\quad
g_{p0}=1-\frac{1}{L_0}\left[\frac{1}{42}+h\left(\frac{u_1^2}{a^2}\right)\right],
\nonumber \\
&& g_p(x)=g_{p0}+\frac{1}{6 L_0}\left[\ln
\left(1+\kappa_1^2\right)+\frac{6 D_{p}\kappa_1^2}
{12+\kappa_1^2}\right],\quad
\nonumber \\
&&  h(z)=-\frac{1}{2}\left[1+(1+z)e^{z}{\rm Ei}(-z) \right],\quad
L_0=\ln(ma)+ \frac{1}{2}-f(Z\alpha),\quad
\nonumber \\
&& a=\frac{111Z^{-1/3}}{m}, \quad f(\xi)=\sum_{n=1}^{\infty}
\frac{\xi^2}{n(n^2+\xi^2)}, \label{8}
\end{eqnarray}
where where $Z$ is the charge of nucleus,  $f(\xi)$ is the Coulomb
correction, the function $g_p(x)$ determines the effective logarithm
using the interpolation procedure, $D_{p}=D_{sc}-10/21=1.8246$,
$D_{sc}=2.3008$ is the constant entering in the radiation spectrum
at $\chi/u \gg 1$ (or in electron spectrum in pair creation process
at $\kappa_1 \gg 1$), see Eq.(7.107) in \cite{BKS},~ Ei($z$) is the
integral exponential function.

The expression for $dW(\omega, y)$ Eq.(\ref{2}) includes both the
coherent and incoherent contributions as well as the influence of
the multiple scattering (the LPM effect) on the pair creation
process (see \cite{BK0}).

3. The expression for the spectral probability of radiation used in
the above derivation can be found from the spectral distribution
Eq.(\ref{2}) ($dW/dy=\omega dW/d\varepsilon $) using the standard
QED substitution rules: $\varepsilon \rightarrow
-\varepsilon,~\omega \rightarrow -\omega,~\varepsilon^2d\varepsilon
\rightarrow \omega^2d\omega$ and exchange $\omega_c(x) \rightarrow
4\varepsilon_c(x)$. As a result one has for the spectral intensity
$dI=\omega dW$
\begin{eqnarray}
&& dI(\varepsilon,y_r)=\frac{\alpha m^2}{2\pi} \frac{y_r
dy_r}{1-y_r} \int\limits_0^{x_0}\frac{dx}{x_0}G_{r}(x, y_r),
\nonumber \\
&&G_{r}(x, y_r)=\int\limits_0^{\infty} F_{r}(x, y_r, t)dt
-r_{3}\frac{\pi}{4},
\nonumber \\
&& F_{r}(x, y_r, t)={\rm Im}\left\lbrace
e^{\varphi_1(t)}\left[r_{2}\nu_{0r}^2 (1+ib_r)f_2(t)+r_{3}f_3(t)
\right] \right\rbrace,\quad b_r=\frac{4\chi^2(x)}{u^2\nu_{0r}^2},
\nonumber \\
&& y_r=\frac{\omega}{\varepsilon}, \quad u=\frac{y_r}{1-y_r},\quad
\varphi_1(t)=(i-1)t+b_r(1+i)(f_2(t)-t), \label{r1}
\end{eqnarray}
where
\begin{eqnarray}
&&r_2=1+(1-y_r)^2,\quad r_3=2(1-y_r),\
\nonumber \\
&&\nu_{0r}^2=\frac{1-y_r}{y_r} \frac{\varepsilon}{\varepsilon_c(x)},
\label{r2}
\end{eqnarray}
where the functions $f_2(t)$ and $f_3(t)$ are defined in
Eq.(\ref{2}). The local value of parameter $\chi(x)$  which
determines the radiation probability in the field Eq.(\ref{4}) is
\begin{equation}
\chi(x)=-\frac{dU(\varrho)}{d\varrho}\frac{\varepsilon}{m^3}=2\chi_s
f(x),\quad \chi_s=\frac{V_0 \varepsilon}{m^3a_s}\equiv
\frac{\varepsilon}{\varepsilon_s},
 \label{r3}
\end{equation}
where $f(x)$ is defined in Eq.(\ref{6}).

The functions and values in Eqs.(\ref{r1}) and (\ref{r2}) (see also
Eqs.(\ref{7}) and (\ref{8})) are
\begin{eqnarray}
&&\varepsilon_c(x)=
\frac{\varepsilon_e(n_a)}{\xi(x)g_r(x)}=\frac{\varepsilon_0}{g_r(x)}e^{x/\eta_1},
\nonumber \\
&&g_r(x)=g_{r0}+\frac{1}{6 L_0}\left[\ln
\left(1+\frac{\chi^2(x)}{u^2}\right)+\frac{6 D_{r}\chi^2(x)}
{12u^2+\chi^2(x)}\right],
\nonumber \\
&&
g_{r0}=1+\frac{1}{L_0}\left[\frac{1}{18}-h\left(\frac{u_1^2}{a^2}\right)\right],\quad
\label{r4}
\end{eqnarray}
where the function $g_r(x)$ determines the effective logarithm using
the interpolation procedure: $D_r=D_{sc}-5/9$=1.7452.

The expression for $dI$ Eq.(\ref{r1}) includes both the coherent and
incoherent contributions as well as the influence of the multiple
scattering (the LPM effect) on the photon emission process (see
\cite{BK1}).

\section{Inclusion of energy loss and photon absorption}

Here we consider the processes of interaction of electrons and
photons with oriented crystal when the target thickness $l$ is of
the order $l\sim L(\varepsilon)$ in the intermediate energy region.

Below we will neglect the energy dispersion (see discussion in
Sec.17.5 \cite{BKS}). On this assumption the energy loss equation
acquires the form
\begin{equation}
dt=\frac{L(\varepsilon)}{\varepsilon}d\varepsilon,\quad
t(\varepsilon,\varepsilon_0)=\int\limits_\varepsilon^{\varepsilon_0}\frac{dx}{x}L(x),\quad
\varepsilon=\varepsilon(\varepsilon_0, t) \label{a1}
\end{equation}
Now the photon spectral distribution taking into account the energy
loss can be written in the form (cp Eq.(20.36) \cite{BKS} for the
total number of photons)
\begin{eqnarray}
&&\omega\frac{dn_{\gamma}^{(in)}}{d\omega}=\int\limits_0^l
\frac{dI(\varepsilon(\varepsilon_0, t), \omega)}{d\omega}
\vartheta(\varepsilon(\varepsilon_0, t)-\omega)dt
\nonumber \\
&&=\int\limits_{\varepsilon_l}^{\varepsilon_0}
\frac{L(\varepsilon)}{\varepsilon}\frac{dI(\varepsilon,
\omega)}{d\omega}\vartheta(\varepsilon-\omega)d\varepsilon,\quad
\varepsilon_l=\varepsilon(\varepsilon_0, l), \label{a2}
\end{eqnarray}
where $dI(\varepsilon, \omega)/d\omega$ is radiation intensity
spectral distribution (see Eq.(\ref{r1})).

For calculation of the photon spectral distribution at the exit from
a target one has to take into account the photon absorption on the
length $l-t$. This can be done by substitution the additional factor
$\exp(-W(\omega)(l-t))$ into the integrand of Eq.(\ref{a2})):
\begin{eqnarray}
&&\frac{dn_{\gamma}^{(out)}}{d\omega}=\frac{1}{\omega}\int\limits_0^l
\exp(-W(\omega)(l-t))\frac{dI(\varepsilon(\varepsilon_0, t),
\omega)}{d\omega} \vartheta(\varepsilon(\varepsilon_0, t)-\omega)dt
\nonumber \\
&&=\frac{\exp(-W(\omega)l)}{\omega}\int\limits_{\varepsilon_l}^{\varepsilon_0}
\frac{L(\varepsilon)}{\varepsilon}\exp(W(\omega)t(\varepsilon,\varepsilon_0))
\frac{dI(\varepsilon,
\omega)}{d\omega}\vartheta(\varepsilon-\omega)d\varepsilon,
\label{a3}
\end{eqnarray}
where $W(\omega)$ is the probability of pair creation by a photon
with the energy $\omega$ per unit time.

In the case $\omega > \varepsilon_l$ the lower limit of the
integrals in Eqs.(\ref{a2}),(\ref{a3}) becomes $\omega$ because of
the function $\vartheta(\varepsilon-\omega)$ in the integrand. In
this case, the integral in Eq.(\ref{a2}) dos't depend on the target
thickness, while the spectral distribution in Eq.(\ref{a3}) contains
the target thickness in the common factor $\exp(-W(\omega)l)$ only.
The difference of the number of the emitted photon and the number of
outgoing photons gives the number of the created
pairs:$~n_p=n_{\gamma}^{(in)}-n_{\gamma}^{(out)}$.

We consider now the hard end ($\varepsilon_0-\omega\ll
\varepsilon_0,~t(\omega, \varepsilon_0)\ll l$) of the spectral
distribution in Eq.(\ref{a3}), where the radiation spectral
intensity is defined by the incoherent contribution only
$(\varepsilon_0-\omega\ll 2\varepsilon_m/3)$. In this situation one
has
\begin{equation}
\frac{dn_{\gamma}^{(out)}}{d\omega}\simeq
\frac{L(\varepsilon_0)g_{r0}}{L_{rad}}\exp(-W(\varepsilon_0)l)
\frac{\varepsilon_0-\omega}{\varepsilon_0^2} \label{a4}
\end{equation}

For soft photons the spectral intensity of radiation depends rather
weakly on the radiating electron energy (see e.g. Fig.3a in
\cite{BK4}). Than using the first equality in Eq.(\ref{a2}) we get
\begin{equation}
\frac{dn_{\gamma}^{(in)}}{d\omega}\simeq
\frac{l}{\omega}\frac{dI(\varepsilon_0, \omega)}{d\omega} \label{a5}
\end{equation}
One can found more accurate expression using the second equality in
Eq.(\ref{a2}) and caring out the averaging procedure given in
\cite{BK1} (see Eqs.(20)-(24)):
\begin{eqnarray}
&&\frac{dn_{\gamma}^{(in)}}{d\omega}=\frac{l}{\omega}\frac{d\tilde{I}}{d\omega},\quad
\frac{d\tilde{I}}{d\omega}=\frac{\tilde{\varepsilon}L(\varepsilon_0)
\frac{dI(\varepsilon_0,\omega)}{d\omega}+
\varepsilon_0L(\tilde{\varepsilon})\frac{dI(\tilde{\varepsilon},\omega)}{d\omega}}
{\tilde{\varepsilon}L(\varepsilon_0)+\varepsilon_0L(\tilde{\varepsilon})},
\nonumber \\
&&
\tilde{\varepsilon}=\varepsilon_0\exp\left(-\frac{l}{\tilde{L}}\right),\quad
\tilde{L}=\frac{\varepsilon_0L(\varepsilon_1)+\varepsilon_1L(\varepsilon_0)}
{\varepsilon_1+\varepsilon_0},\quad
\varepsilon_1=\varepsilon_0\exp\left(-\frac{l}{L(\varepsilon_0)}\right).
\label{a3a}
\end{eqnarray}

The probability of soft photons absorption is determined by the
incoherent contribution of pair creation $W(\omega)\simeq
7g_{p0}/9L_{rad}$, where $L_{rad}$ is the radiation length in a
corresponding amorphous medium, and is small for the target
thickness under consideration $l\sim L\ll L_{rad}$. In this case
using the difference of the first equalities in Eq.(\ref{a2}) and
Eq.(\ref{a3}) we find for the spectral distribution of created pairs
$\omega=\varepsilon^+ +\varepsilon^-$
\begin{equation}
\frac{dn_{\gamma}^{(in)}}{d\omega}\simeq
\frac{l^2}{2\omega}\frac{dI(\varepsilon_0,
\omega)}{d\omega}W(\omega) \label{a6}
\end{equation}

In the process of pair photoproduction the spectral distribution of
created particles changes due to the radiation energy loss before
leaving a target. Let us introduce the integral with
$\delta$-function into the initial distribution. We have at the
distance $t$ before leaving a target
\begin{equation}
dw(\varepsilon, \varepsilon', t)=-\frac{dW(\omega,
\varepsilon')}{d\varepsilon'}
\exp(-W(\omega)(l-t))\delta(t(\varepsilon, \varepsilon')-t)
\frac{dt(\varepsilon, \varepsilon')}{d\varepsilon}d\varepsilon
d\varepsilon'dt, \label{a7}
\end{equation}
where $dW(\omega, \varepsilon')/d\varepsilon'$ is the spectral
distribution over energy of one of the created particles in the
point of creation (see Eq.(\ref{2})) at the distance $t$ before
leaving a target, the function $t(\varepsilon, \varepsilon')$ is
defined in Eq.(\ref{a1})). The factor $\exp(-W(\omega)(l-t))$ is the
probability of the initial photon to survive at the depth $l-t$.
Caring out the integration in Eq.(\ref{a7})) over variables
$\varepsilon'$ and $t$ we get
\begin{equation}
\frac{dw(\omega,
\varepsilon)}{d\varepsilon}=\frac{L(\varepsilon)}{\varepsilon}
\exp(-W(\omega)l)\int\limits_{\varepsilon}^{\varepsilon_p}
\frac{dW(\omega,
\varepsilon')}{d\varepsilon'}\exp(W(\omega)t(\varepsilon,
\varepsilon'))\vartheta(\omega-\varepsilon') d\varepsilon',
\label{a8}
\end{equation}

In the case $\omega < \varepsilon_p$ the upper limit of the integral
in Eq.(\ref{a8}) becomes $\omega$ because of the function
$\vartheta(\omega-\varepsilon')$ in the integrand. In this case, the
integral in Eq.(\ref{a8}) dos't depend on the target thickness,
which enters in the external factor $\exp(-W(\omega)l)$ only.

Let us consider the hard end ($\omega-\varepsilon\ll
\omega,~t(\varepsilon, \varepsilon')\ll l$) of the spectral
distribution in Eq.(\ref{a8}), where the spectral probability of
pair creation is defined by the incoherent contribution only
$(\omega-\varepsilon\ll 2\omega_m/3)$. In this situation one has
\begin{equation}
\frac{dw(\omega, \varepsilon)}{d\varepsilon}\simeq
\frac{L(\omega)g_{p0}}{L_{rad}}\exp(-W(\omega) l) \frac{\omega
-\varepsilon}{\omega^2} \label{a9}
\end{equation}
This formula has the same structure as Eq.(\ref{a8}).

In Fig.2 the spectral distribution of radiation $\omega
dn_{\gamma}^{(in)}/d\omega$ is shown.  The final electron energies
are correspondingly $\varepsilon_l=10~$GeV and
$\varepsilon_l=16~$GeV for the used thicknesses. It is seen that at
$\omega >16~$GeV the curves 2 and 4 merge. For $\varepsilon_0-\omega
\ll 5~$GeV the spectral distribution is in a good agreement with
Eq.(\ref{a4})(without the factor $\exp(-W(\varepsilon_0)l)$ ) (at
$\omega>19~$GeV it is better than 5\%). This factor, considering the
photon absorption with the energy $\omega \simeq \varepsilon_0$ when
photon is crossing the whole crystal, is 0.85 and 0.91 for the used
thicknesses. The curve 2 and 4 calculated for thin targets differ
only on a scale (ratio of their ordinates coincides with the ratio
of thicknesses). In the soft part of spectrum the difference between
the curves 1 and 2 is not very large. It is in agreement with
Eq.(\ref{a5}). This property of spectra in oriented crystals was
indicated in \cite{BK}. In the hard part of spectrum the difference
is quite essential and one has to take into account the energy loss
in the case where the crystal thickness $l \simeq L(\varepsilon)$.
In the case $l \ll L(\varepsilon)$ the essential distortion of the
spectral curve occurs in the hard end of spectrum only as it is seen
from comparison of the curves 3 and 4.

In Fig.3 the spectral distribution of radiation $\omega
dn_{\gamma}^{(in)}/d\omega$  is shown. These parameters are used in
the experiment NA63 carried out recently at SPS at CERN (for
proposal see \cite{NA63}). The final electron energies are
correspondingly $\varepsilon_l=124~$GeV and $\varepsilon_l=153~$GeV
for the used thicknesses. It is seen that at $\omega >150~$GeV the
curves 2 and 4 merge. For $\varepsilon_0-\omega \ll 60~$GeV the
spectral distribution is in a good agreement with
Eq.(\ref{a4})(without the factor $\exp(-W(\varepsilon_0)l)$ ) (at
$\omega>170~$GeV it is better than 8\%). This factor, considering
the photon absorption with the energy $\omega \simeq \varepsilon_0$
when photon crossing the whole crystal, is 0.94 and 0.97 for the
used thicknesses. The curve 2 and 4 calculated for thin targets
differ only on a scale (ratio of their ordinates coincides with the
ratio of thicknesses). In the soft part of spectrum the difference
between the curves 1 and 2 is not very large. It is in agreement
with Eq.(\ref{a5}).  In the hard part of spectrum the difference is
quite essential and one has to take into account the energy loss in
the case where the crystal thickness $l \simeq L(\varepsilon)$. In
the case $l \ll L(\varepsilon)$ the essential distortion of the
spectral curve occurs in the hard end of spectrum only as it is seen
from comparison of the curves 3 and 4.

In Fig.4 the spectral distribution of radiation $\omega
dn_{\gamma}^{(out)}/d\omega$ in a quite thick target is shown.  The
final electron energy at the exit from the target is
$\varepsilon_l=8.1~$GeV. The factor, characterizing the additional
suppression of the spectral distribution in the hard end, is
$\exp(-W(\varepsilon_0)l)$=1/4.5. The quite substantial difference
between the solid and dashed curves is connected with the relatively
large number of $e^+e^-$ pairs created by radiated photons. The
particles of these pairs emit photons also. However, a consideration
of the next stage of cascade is out of scope this paper.

In Fig.5 the distortion of the positron spectrum is shown due to the
photon emission from positrons and the initial photon absorption in
a rather thick target. The curves 3 and 4 calculated for the thin
target are symmetric with respect to point $\omega$/2=10~GeV. The
area under the curves 1 and 2 divided by $\omega=20~$GeV give the
total number of created pairs in the interval from 1 to 19 GeV's.
Since at calculation of the curves 3 and 4 the photon absorption was
not taken into account the area under these curves is by 10\%(5\%)
larger than under the curves 1 and 2 respectively. The hard end of
the curves 1 and 2 is in a good agreement with Eq.(\ref{a9}).
According to this equation the difference between these curves is
due to difference of the factors $\exp(-W l)$ equal to 0.8 and 0.9
respectively.

\section{Conclusion}

In this paper the spectral distribution of particles of
electron-positron pair created by a photon and the spectral
distribution of radiation from an electron in an oriented crystal of
the intermediate thickness $l\sim L(\varepsilon_0)$ is calculated
for the energy $\varepsilon(\omega)\sim \varepsilon_m$. The theory
approach is developed which takes into account the energy loss due
to the photons emission from charged particles as well as the
absorption of photons in a crystal. For the first time the closed
analytical expressions were obtained for description of the starting
stage of the electron-photon cascade in oriented crystal. Since in
the energy region under consideration all the specific mechanisms
(coherent and incoherent radiation(pair creation), the LPM-effect)
are essential, we used the recently developed method which includes
all these mechanisms (see Eqs.(\ref{2}), (\ref{r1})). It turn out to
be important at the cascade analysis that for the energies and
thicknesses under consideration the probability of secondary
processes is suppressed because of the relatively large photon
absorption length $L_{pr}(\omega\sim \varepsilon_m) \gg
L(\varepsilon\sim \varepsilon_m)$. Some estimates of the number of
electrons and photons at the exit of target are given Sec.20.4
\cite{BKS}, where one can found also the results of simulation of
experimental data.

The results obtained show the substantial variation of the hard
photon spectral distribution comparing with thin target even for the
relatively thin targets (see Figs.2,3). At the same time the
deformation of the soft part of spectra is quite modest. The found
positron energy distribution at $l\sim L(\varepsilon)$ demonstrates
the dramatic difference from the spectrum in thin crystal. In the
relatively thick tungsten crystal  $l\gg L(\varepsilon_0)$ (Fig.5)
and at the high electron energy $\varepsilon_0\gg \varepsilon_m$ we
found the dramatic variation of the emitted photon spectrum
comparing with thin crystal. In this case the appreciable number of
secondary pairs are created and the next stages of cascade process
should be studied.

\vskip3mm

{\bf Acknowledgments}

The authors are indebted to the Russian Foundation for Basic
Research supported in part this research by Grant 06-02-16226.

\newpage

{\bf Figure captions}

{\bf Fig.1} The radiation length
$L(\varepsilon)=\varepsilon/I(\varepsilon)$ (curve 1)and the pair
creation length $L_{pr}(\omega)=1/W(\omega)$ (curves 3) in the
tungsten crystal, axis $<111>$, temperatures T=100 K. The values
$L(\varepsilon)$ (curve 2) and $L_{pr}(\omega)$ (curve 4) in the
germanium crystal, axis $<110>$, temperatures T=293 K  vs the
electron ($\varepsilon$) or photon ($\omega$) energy.

{\bf Fig.2}

The spectral distribution of radiation at the initial electron
energy $\varepsilon_0=20~$GeV in the tungsten crystal, axis $<111>$,
T=100 K in two targets with thickness
$l=0.032~$cm$=0.77~L=0.16~L_{pr}$ (curves 1 and 2) and
$l=0.01~$cm$=0.24~L=0.093~L_{pr}$ (curves 3 and 4) vs the photon
energy $\omega$. The curves 2 and 4 are calculated according to
Eq.(\ref{r1}), while the curves 1 and 3 are calculated according to
Eq.(\ref{a2}) which takes into account the electron energy loss.

{\bf Fig.3}

The spectral distribution of radiation at the electron with energy
$\varepsilon_0=180~$GeV in germanium crystal, axis $<110>$, T=293 K
in two targets with thickness $l=0.04~$cm$=0.38~L=0.064~L_{pr}$
(curves 1 and 2) and $l=0.017~$cm$=0.16~L=0.027~L_{pr}$ (curves 3
and 4) vs the photon energy $\omega$. The curves 2 and 4 are
calculated according to Eq.(\ref{r1}), while the curves 1 and 3 are
calculated according to Eq.(\ref{a2}) which takes into account the
electron energy loss.

{\bf Fig.4}

The spectral distribution of radiation at the electron initial
energy $\varepsilon_0=100~$GeV in the tungsten crystal, axis
$<111>$, T=100 K in the target with thickness
$l=0.1~$cm$=3.15~L(\varepsilon_0)=1.505~L_{pr}(\varepsilon_0)$. The
dotted curve is calculated according to Eq.(\ref{r1}), the dashed
curve is calculated according to Eq.(\ref{a2}) which takes into
account the electron energy loss and the solid curve is calculated
according to Eq.(\ref{a3}) which takes into account both the
electron energy loss and the photon absorption,

{\bf Fig.5}

Spectra of positrons created by the photon with energy
$\omega=20~$GeV in the tungsten crystals (axis $<111>$, T=100 K)
with thickness $l=0,0414~$cm=$L(\omega)$ and
$l=0,0207~$cm=$L(\omega)$/2. The curves 3 and 4 are calculated for
thin targets according to Eq.(\ref{2}), while the curves 1 and 2 are
calculated according to Eq.(\ref{a8}) which takes into account both
the positron energy loss and the photon absorption,

\newpage
\begin{table}
\begin{center}
{\sc Table}~ {Parameters of the pair photoproduction and radiation
processes in the tungsten crystal, axis $<111>$ and the germanium
crystal, axis $<110>$ for two temperatures T
($\varepsilon_0=\omega_0/4, \varepsilon_m=\omega_m,
\varepsilon_s=\omega_s$)}
\end{center}
\begin{center}
\begin{tabular}{*{10}{|c}|}
\hline Crystal& T(K)&$V_0$(eV)&$x_0$&$\eta_1$&$\eta$&
$\omega_0$(GeV)&$\varepsilon_m$(GeV)&$\varepsilon_s$(GeV)&$h$ \\
\hline W & 293&417&39.7&0.108&0.115&29.7&14.35&34.8&0.348\\
\hline W &100&355&35.7&0.0401&0.0313&12.25&8.10&43.1&0.612\\
\hline Ge & 293 & 110& 15.5
&0.125&0.119&592&88.4&210&0.235\\
\hline Ge & 100 & 114.5& 19.8
&0.064&0.0633&236&50.5&179&0.459\\
\hline
\end{tabular}
\end{center}
\end{table}

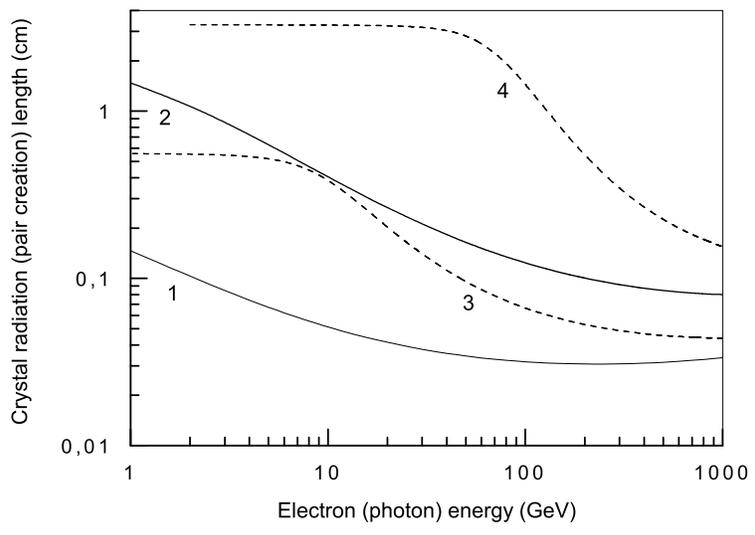
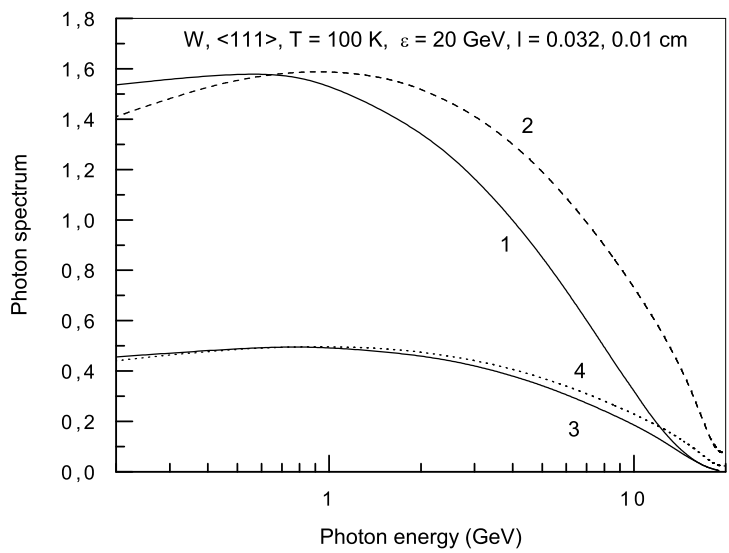
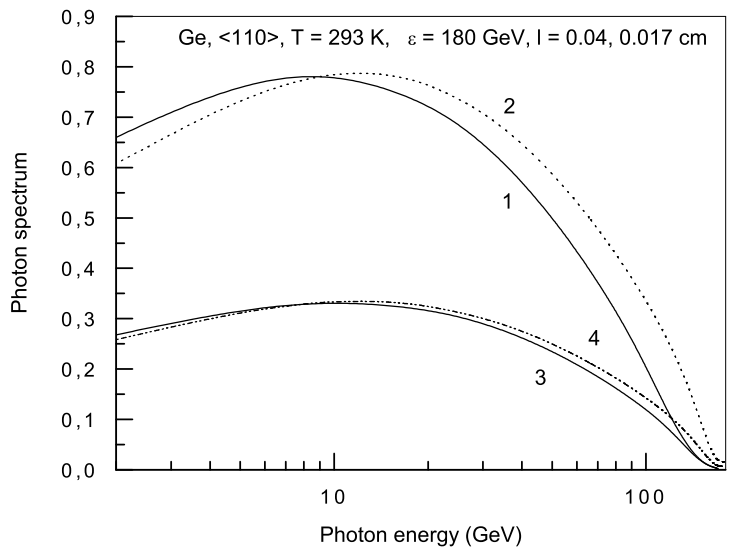
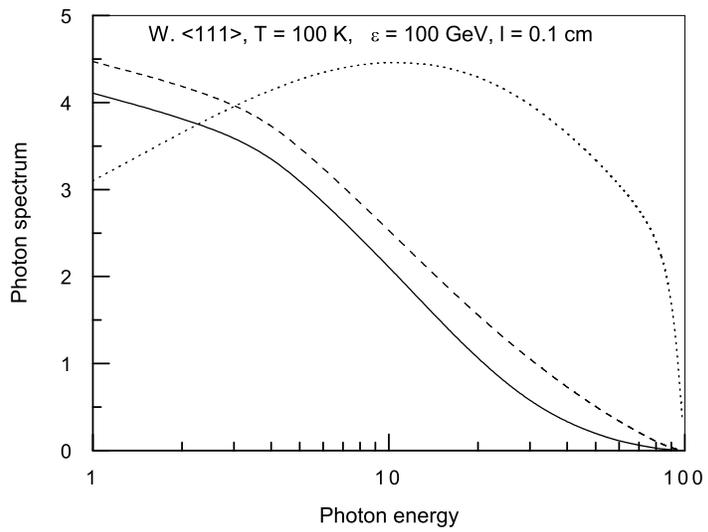
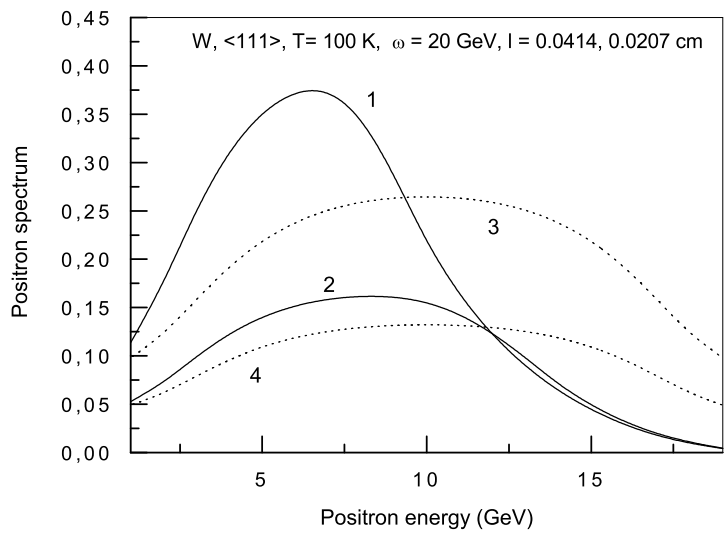
\end{document}